# An Intuitive Tangible Game Controller

Jacques Foottit, Dave Brown, Stefan Marks & Andy M. Connor
Auckland University of Technology
Colab (D-60), Private Bag 92006
Wellesley Street, Auckland 1142, NZ
+64 9 921 9999

[wgk9328; wpv9142; smarks; aconnor]@aut.ac.nz

## ABSTRACT
This paper outlines the development of a sensory feedback device providing a tangible interface for controlling digital environments, in this example a flight simulator, where the intention for the device is that it is relatively low cost, versatile and intuitive. Gesture based input allows for a more immersive experience, so rather than making the user feel like they are controlling an aircraft the intuitive interface allows the user to *become* the aircraft that is controlled by the movements of the user's hand. The movements are designed to allow a sense of immersion that would be difficult to achieve with an alternative interface.

A vibrotactile based haptic feedback is incorporated in the device to further enhance the connection between the user and the game environment by providing immediate confirmation of game events. When used for navigating an aircraft simulator, this device invites playful action and thrill. It bridges new territory on portable, low cost solutions for haptic devices in gaming contexts.

## Categories and Subject Descriptors
H.5.2 [**User Interfaces**]: Haptic I/O

## 1. INTRODUCTION
Haptic technology [1] has found acceptance across a broad spectrum of fields, from highly sophisticated simulators for training surgeons to everyday devices like mobile phones. The diversity of applications also means there is a great diversity in the forms of implementation. Haptic feedback technology can be broadly divided into two main categories. The first category utilise force reflecting interfaces which are capable of providing highly realistic representation of real-world experiences, as they apply constant forces to the user based on calculated forces from the simulation. However this fidelity comes at a cost, and often these interfaces are large, heavy and power consuming. The second category for haptic feedback technology is abstracted simulations where the haptic feedback provides information to the user that is not a literal representation of real-world forces. For example haptic feedback can be used to convey emotion [2], or to draw the attention of a user to a particular output. A wide variety of implementations exist and the focus of this paper is the use of a vibrotactile feedback mechanism.

This paper outlines the development of a tangible interface for controlling digital environments such as games that has the potential for being versatile and intuitive. The interface is based around a particular design of a haptic glove used in conjunction with a Kinect controller. Most haptic devices today are large, clunky, mechanical exoskeletal devices, connected to external power sources. The motivation for the interface described in this paper is the provision of a low cost, light weight and wearable device. The haptic glove uses a range of sensors to identify hand movements and vibratory motors as actuators that are mounted in the fingertips and are triggered by events in the game environment. The main difference with other wearable haptic devices is that the glove can also trigger game events through a combination of hand position tracking and other on-board sensors.

## 2. BACKGROUND & RELATED WORK
Humans are social creatures and as such find touch, physical interaction and human-to-human presence essential for the enjoyment of life [3]. Computer entertainment can also provide humans with enjoyment, by allowing virtual fantasy and imaginative play activity to be carried out. However, present computer based entertainment is often limited by the lack of tangible interaction. Whilst game controllers have advanced since the days of "keyboard bashing" in the likes of the popular 1984 Daley Thomson's Decathlon game, almost all modern game controllers can be considered as devices external to the player and as such limit the extent to which a player can immerse themselves in an intuitive way.

Recent research suggests that the degree of body movement imposed, or allowed, by a game controller directly impacts the engagement levels of a player and that an increased involvement of the body can afford the player a stronger affective experience [4, 5]. Controller-less gameplay can be facilitated through the use of technology like the Kinect controller, and this has the potential for smooth and natural user interfaces [6]. However, such intuitive gameplay is in a way still intangible due to a lack of physical connection with the game environment, in particular the lack of any form of feedback that results from game events.

### 2.1 Tangible Interaction
Research in Tangible Interaction has been inspired by many different disciplines, including psychology, sociology, engineering and human-computer interaction (HCI). Tangible Interaction encompasses user interfaces and interaction approaches that emphasise the tangibility and materiality of the interface, physical embodiment of data, whole-body interaction and the embedding of the interface and the users' interaction in real spaces and contexts [7].

There have been significant developments in terms of tangible game interfaces where players interact with the game environment by physically grasping and moving real-world objects [8, 9], where real world games are augmented and enhanced using technology [10, 11] and even hybrid approaches [12]. It has been found by some researchers that tangible interfaces produce more intuitive interaction [13] that require less instruction for use. It has also been observed that real body motions projected into a virtual environment games can be deciphered by the player as relating to themselves [14]. However, even such complex tangible interfaces lack any feedback mechanism and it is argued that a haptic tangible device may further enhance the game play experience by creating a conduit for such feedback to take place.

Most tangible interfaces are purposefully designed as external objects through which a player interacts. This project has adopted an alternative approach which is to consider how an interface can be considered more of an integral part of the human body.

## 2.2 Actor-Network Theory

The Actor-Network Theory as proposed by Latour [15] is a relatively new theory in the field of sociology. The core of this new theory is the attempt to combine the idea of technological networks and social networks into a unified system. This arises from the observation that the technical and the social cannot truly be viewed in isolation. As a result, technological objects become part of the same network as people, along with virtually all other objects.

In general, technology is created to support and interact with humans, and at the same time new technologies are shaping the way that humans interact with each other. Latour is not alone in his view of technology either. Amber Case presents a similar view when she says "The most successful technology gets out of the way and helps us live our lives. And really, it ends up being more human than technology, because we're co-creating each other all the time." [16]. Features like Siri, Apple's virtual assistant, epitomise this underlying goal of creating more human technology. Here the interaction is intentionally designed to be more than just conveying information. The virtual interface is given a personality, and speaks with language that we associate with other humans more than with computers. At the same time, technology is becoming even more integrated into the human experience. Technologies like the mobile phone have not only radically changed the way people communicate but increasingly are becoming more like augmentations, blurring the line between the human actor and the machine.

Whilst some researchers have considered the value of Actor-Network Theory in the design of games and virtual environments [17] there has been little work to date that explicitly considers the role of game controllers as being integral to the player rather than an interface between the player and the game. The goal of this project is to create a wearable interface that seamlessly extends the natural movements of the human body into an intuitive game interface.

## 2.3 Wearable Haptic Devices

A full review of related work is beyond the scope of this paper, but the development of the glove has been informed by a rigorous systematic review of recent literature [18]. This review identified a number of devices under development that utilised different mechanisms of haptic feedback, including vibrotactile feedback. Many of these devices rely purely on vibrotactile feedback, though in some cases this was combined with other forms of haptic feedback including heat [19], pressure [20], and electrical muscle stimulation [21].

Vibrotactile feedback was used in a wide variety of applications compared to force feedback, and also was mostly used to portray abstracted information rather than literal simulations of touch. For example, one of the devices was designed to vibrate when a child with ADHD lost focus on a task and in order to restore their attention[22]. Two were designed to portray visual information to a visually impaired user, e.g., by utilising vibration to portray distances from objects [23], or to alert the user to obstacles [24].

The devices utilising vibration also varied in position on the body, including the torso [19, 23, 25], arms [20-22] and head [24]. In contrast, the review identified that only the devices utilising force feedback were mounted on the hand. One of the devices utilised a custom designed actuator to produce vibration [26], with the remaining devices mostly utilising off the shelf components. This included a vibrotactile glove that exclusively utilised off the shelf components [27].

Much of the work in this area is of a "serious" nature and doesn't emphasise the potential for more fun based activities. There is a potential to develop a low cost, wearable haptic game controller as the first step towards more immersive and tangible interfaces to a variety of digital environments. There is an emerging body of work that considers the integration of motion controllers, such as the Kinect, with haptic feedback. For example, Frati & Prattichizzo have developed a system that provides simple, low cost haptic feedback to the fingertips by integrating hand tracking using a Kinect with vibrotactile feedback [28]. This approach only provides feedback and does not allow the user to actively interact with the virtual environment, for example the triggering of game events. The inclusion of this bi-directional exchange of information between the player and the game is the major contribution of the current work.

## 3. THE HAPTIC GLOVE

Throughout the development of the glove we explored how to integrate technology into the human experience. It is a curious thing when a device becomes so natural that it is almost like an extension of the person. Whilst there are many advances in user interfaces that aim towards making the technology more natural and intuitive, achieving a truly integrated experience where the interface becomes part of a person's experience of their self remains rare. At the forefront of this integration of technology and humanity is prosthetics and orthotics; technological devices designed specifically to integrate with the human body. Devices like mobile phones and cars that radically transform the way a person can interact with the world around them also tend to become integrated into the human experience over time, although these devices are much less likely to become a part of the user's perception of their self the way an orthotic or prosthetic might.

A common thread emerges when looking at the technologies that successfully integrate into the human experience – the technologies must fit well with the human body. In our project, it quickly became clear that it would be vital for our glove to fit the hand comfortably and be light enough not to impede the mobility of the hand. Wireless communication was also important for our project, as being tethered to a computer creates a physical and psychological barrier that separates the technological device from the user's perception of their self. A great deal of this change in the experience came from the need to keep track of the cable when using the wired solution. It distracted from the user experience by requiring them to be aware of the position of the cable in order to avoid it getting tangled or pulled out.

The key to wearable technology is effective miniaturization without losing features associated with larger devices. As devices become smaller and more energy efficient, it becomes possible to embed them into worn artefacts. In the case of providing input to a computer, a particularly useful piece of technology that has developed greatly in recent years is the Micro-Electro-Mechanical Systems (MEMS) based Inertial Measurement Unit (IMU). These remarkable units allow for the combination of sensors such as accelerometers, gyroscopes and magnetometers into incredibly small form factors and were a key aspect of the final glove design. In fact, these devices can be as small as a few millimetres in length, width and height. They are also becoming increasingly affordable, and have become ubiquitous in mobile technology such as smart phones.

Although there has been much improvement in this area, these small, affordable devices are still considered somewhat inaccurate compared to their more expensive and bulky counterparts. However their accuracy is sufficient for most consumer applications, making them an ideal solution for providing orientation information for a glove-based input device. Their key limitation is a tendency to drift over time – particularly when being used to provide positional information. To overcome this limitation, technologies like the Microsoft Kinect can be used to supplement the data provided by these devices [26].

The development of the glove utilised a rapid prototyping methodology with various features trialled and refined to produce the final design[1]. In particular, the prototyping involved considerable experimentation with different types of glove fabric, sensor and mounting for the haptic and other electronic components. One of the early test prototypes is shown in Figure 1.

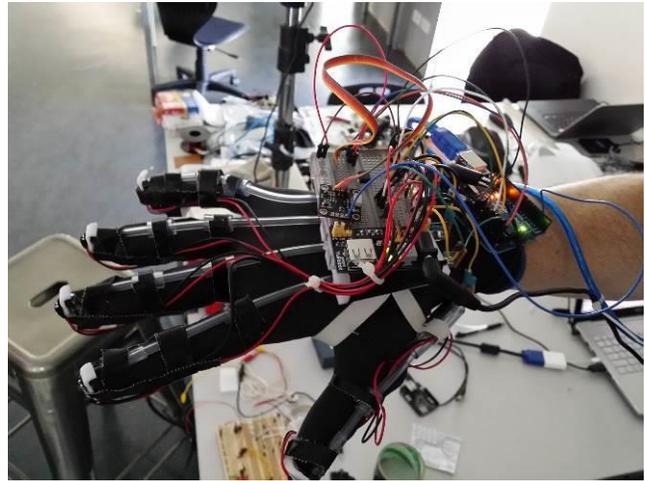

**Figure 1. Initial haptic glove prototype.**

The flexibility, lightness and small size of the glove become an immediate focus, and using a custom fabricated PCB and a small form factor Arduino contributed a great deal towards achieving our goals in this area. The use of a custom knitted glove was also significant, as it was far more comfortable than the early prototypes. This was primarily due to the flexible nature of the fabric that still held the optical sensors in place. Again this was an iterated process and the comfort of the glove was a prime consideration during development. The final glove is shown in Figure 2.

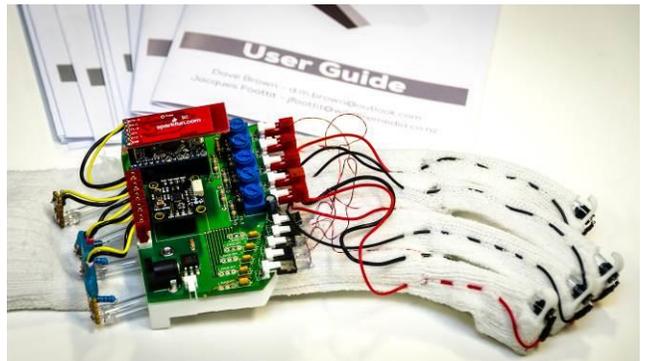

**Figure 2. The final haptic glove.**

This project required a wide range of support from different disciplines. The disciplines ranged from engineering to fashion to health sciences. This required interaction and coordination with people with very different sets of knowledge, each of which had something to contribute to the project. The role of the project team was to integrate the different aspects into a unified design, and to help each of the supporting people to understand enough of the other parts of the project to provide useful input. This bridging of disciplines to achieve a goal that could not be achieved without an understanding of the importance of communication in transdisciplinary design projects. Each member of the design team had their own leanings and preferences towards certain aspects of the project, but the greatest skills developed throughout this project were the ability to draw on the expertise of others to support the goals for the project.

---

[1] Development show reel: http://vimeo.com/98431538

## 4. GAME ENVIRONMENT & CONTROL

The potential for the haptic glove to operate as an effective and intuitive tangible interface has been demonstrated through the development of a basic flight simulator in Unity that is controlled by the movements of the user's hand. The required movements are designed to feel intuitive and allow for a sense of immersion that would be difficult to achieve with an alternative interface. In this example the user's hand can *become* the aircraft much the same way that a child would imagine it.

The flight simulator is controlled through a combination of the haptic glove and a Microsoft Kinect. The glove utilises a flex sensor for each finger and an IMU feeding into Unity to control virtual elements. The player moves their hand closer or further away from the Kinect to change the velocity of the aircraft with hand orientation changing aircraft placement. To fire guns the player has the choice of bending the thumb or clenching the fist to shoot missiles. The vibratory motors on the fingertips cause haptic responses to the player. The controls excel at generating a learning basis for finger movement and placement, intuitively forcing creative hand gestures and participation. Figure 3 illustrates the control mechanism implemented by the flight simulator.

The game itself is rather limited, whilst the player can control their own aircraft and attack enemy vessels, these vessels are essentially static entities that do not fight back. As it was implemented, the game also had no capacity for a player to crash into objects, the ground or to lose a life. Whilst perfectly functional as a demonstration of the haptic glove, the game would have little appeal as a game in its own right and would likely have limited ability to engage a player for more than a few minutes. The significance of this will be discussed in Section 5.

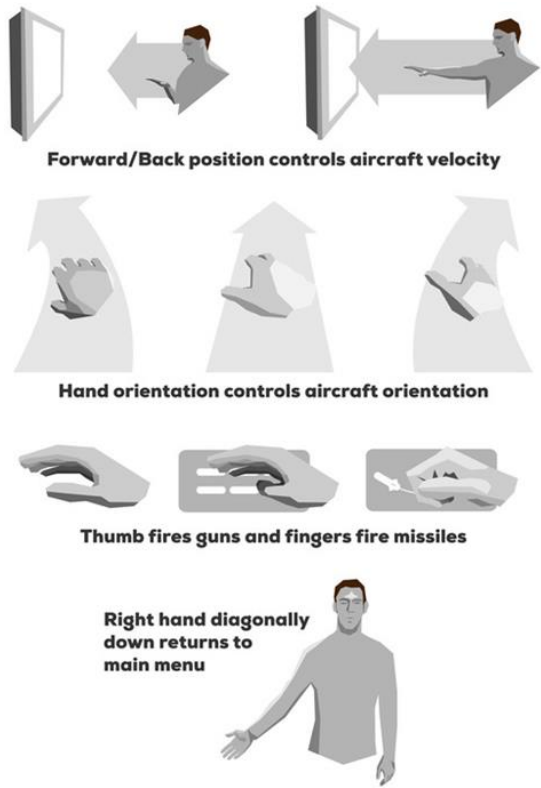

**Figure 3. Game control instructions.**

## 5. EVALUATION

A formal usability study of the haptic glove has not yet been conducted. The glove is currently undergoing considerable refinement and formal evaluation will be postponed until this is completed. However, it is possible to draw anecdotal conclusions regarding the usage of the glove from the extensive play testing conducted during development and also from observations and discussions with users who trialled the glove as a game interface at a public demonstration.

Figures 4 and 5 show players using the glove as a game interface at this demonstration.

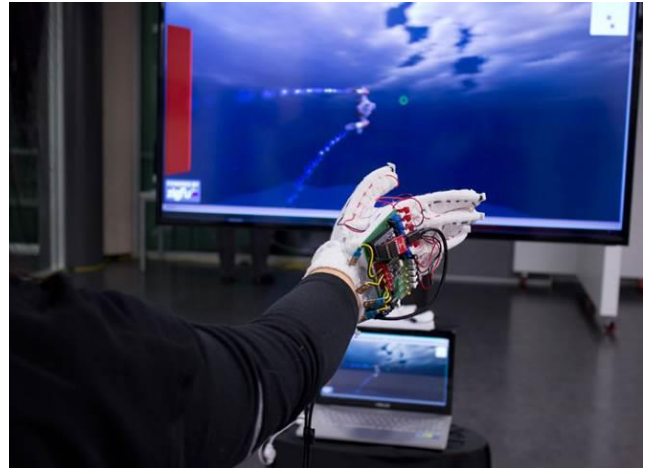

**Figure 4. The haptic glove in action at a public demonstration.**

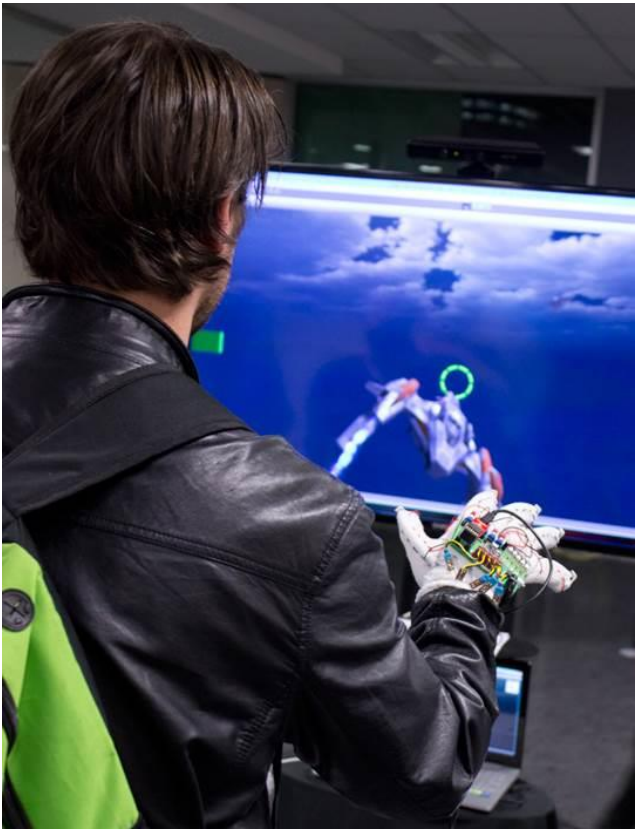

**Figure 5. The haptic glove in action at a public demonstration.**

The glove was demonstrated alongside a broad range of other student work during a public open exhibition day at the end of the semester, including a number of other game related projects. It was observed that the project attracted considerable attention with many people wanting to try the game and the glove. Of particular interest was the degree of engagement that players had with the game, with many players not wanting to stop playing. This is quite remarkable considering the limited functionality of the game as outlined in Section 4. In addition to the time spent playing the game, it was interesting to observe the behaviour of the players which in many cases was playful, light hearted and frivolous. The glove was not really considered to be an interface or an external object, instead it promoted movement and interaction in a much more childlike and natural way.

Most people quickly adapted to the use of the glove in the game without referring to the usage instructions, suggesting that the interface and controls are suitably intuitive. However, it should be considered that the exhibition event did attract a particular "techno-curious" type of person. It is therefore important to not overstate observations regarding intuitiveness until a more formal usability study is undertaken. However, in general, it was noted that the control of the orientation of the aircraft was mastered more quickly than the control of the aircraft velocity. The Kinect had some trouble tracking the right person due to the number of people watching, but this didn't impact the fun very much as it only influenced the speed.

Verbal feedback from the users of the glove was extremely positive, with almost everyone commenting that the haptic sensations were effective and useful prompts for appreciating the action in the gameplay. Comments were also received about the intuitive nature of the controls that reinforced our observations on the night as well as our experiences encountered whilst play testing throughout the development of the glove. This was our primary aim for the glove, so to hear it from users outside of the development team confirmed that the glove had at least gone some way towards achieving this goal.

## 6. FUTURE WORK

Whilst the developed prototype has been productive in terms of developing a two way tangible interface, there is still considerable scope to extend this work in new directions as well as focusing on refining the initial prototype. Questions have arisen through the development of the glove around the nature of immersion and engagement in virtual environments.

Games are just one example of a virtual environment, however most games are still limited to desktop monitors where as other virtual environments have been extended into the physical space immediately surrounding the user. Whilst some researchers argue that gloves are unwieldy in terms of tracking gestures in virtual environments [29] this ignores the potential for the haptic feedback. Work is currently in progress that uses wide area motion capture technology in conjunction with an Oculus Rift to allow the interaction with large scale data visualisation [30].

One specific example is the visualisation of the neurons in a spatio-temporal artificial neural network, where a motion capture "mouse" can be used to select individual neurons. This experience can easily be augmented through the application of a haptic glove as can many others. Figure 6 shows the glove being used in an environment where the user can interact with a number of virtual objects. Whilst this system is operational it has as yet not undergone any form of evaluation other than informal testing during the development of the glove.

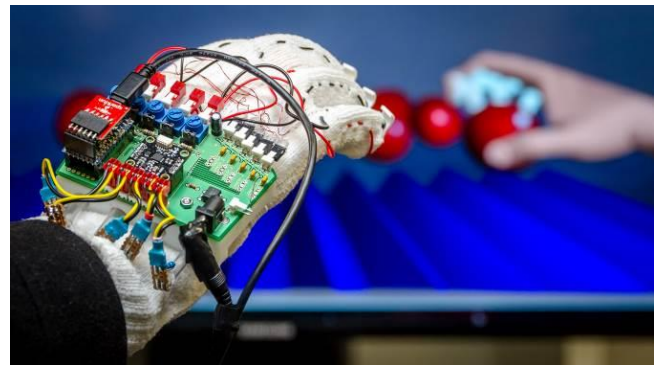

**Figure 6. Object interaction simulator.**

The integration of the object interaction simulator with wide area motion capture would allow studies to be undertaken related to engagement and immersion. However, there is still potential to revisit the role of games and play in such situations. There are a number of open questions about how to design and implement games that free the player from the constraints of typical desktop monitors.

Whilst such open ended questions provide a number of longer term goals, there are more immediate and practical areas for consideration. For example, the glove itself is currently dependent on the Kinect for external motion tracking and this is a limitation in a number of ways, not least of which is lack of fidelity in motion tracking in busy environments.

Work is currently in progress to implement is a translation from wrist orientation to screen position so that the glove could act like a mouse input without the need for motion tracking. However, such approaches only provide a degree of orientation data and will not necessarily be able to accurately pinpoint the device in a three dimensional space envelope. An important question for further work is whether such positional accuracy can be achieved without the need for external tracking.

One of the main limitations of the current work is the relative simple, holistic evaluation of the system. Therefore, the main direction for future work will entail a thorough and systematic usability evaluation of the glove. This will be undertaken when current revisions to the first prototype are complete and will entail not just investigating usability and engagement in the context of a use as a games controller but will also consider other types of digital and virtual environments. This evaluation will allow the suitability of the glove as a generic haptic controller to be evaluated.

## 7. CONCLUSIONS

The haptic feedback glove project has explored an alternative method of interaction with machines by creating an artefact capable of providing both gesture based input and haptic feedback. The potential applications of such technologies are widespread and not limited to simple game controllers. The emphasis of the project was to create a device that seamlessly bridges from the user experience "in the real world" to the actions in the virtual world. It challenges the notion of what a tangible interface is by becoming an invisible mediator between the human and the machine, yet one that provides physical engagement potential. This enables a more intuitive and immersive experience for the user. The haptic feedback in particular is vibration based rather than force reflecting, allowing for low power and portability but making it impossible to restrict user motion based on virtual stimuli.

As this technology develops further, the interactions between humans and machines will be able to more closely resemble the interactions between humans and each other. This has the potential to improve the immersion of virtual experiences, as well as reducing errors and improving input speed for a range of tasks. There is even the potential for this technology to allow for new methods of learning that are yet to be discovered.